\documentclass[twocolumn,shortnote]{jpsj3}
\usepackage{txfonts}
\usepackage{epstopdf}

\title{Bulk Superconductivity in Fe$_{1+y}$Te$_{1-x}$Se$_{x}$ Induced by Annealing in Se and S Vapor }

\author{Yue Sun$^{1,2}$, Yuji Tsuchiya$^2$, Tatsuhiro Yamada$^2$, Toshihiro Taen$^2$, Sunseng Pyon$^2$, Zhixiang Shi$^{1*}$, and Tsuyoshi Tamegai$^{2\dag}$}
\inst{$^1$Department of Physics, Southeast University, Nanjing 211189, People's Republic of China \\
$^2$Department of Applied Physics, The University of Tokyo, 7-3-1 Hongo, Bunkyo-ku, Tokyo 113-8656, Japan} 

\abst{}

\kword{Fe$_{1+y}$Te$_{1-x}$Se$_{x}$, Se and S annealing effect, iron chalcogenide, excess Fe, critical current density}

\begin{document}
\maketitle
Stimulated by the discovery of high-temperature superconductivity in LaFeAs(O,F),\cite{1} a series of iron-based superconductors (IBSs) have been discovered.\cite{2} Among these, FeTe$_{1-x}$Se$_{x}$ composed of only Fe(Te,Se) layers has attracted special attention due to its simple crystal structure, which is preferable for probing the mechanism of superconductivity. On the other hand, its less toxic nature makes FeTe$_{1-x}$Se$_{x}$ a more suitable candidate for applications among the family of IBSs.\cite{3} However, superconductivity and magnetism of this compound are not only dependent on the Se doping level, but also sensitively on Fe non-stoichiometry, which originates from the partial occupation of excess Fe at the interstitial site in the Te/Se layer.\cite{4,5} The excess Fe with valence Fe$^+$ will provide an electron into the system, and it is also strongly magnetic, which provides local moments that interact with the adjacent Fe layers.\cite{6} The magnetic moment from excess Fe will act as a paring breaker and also localize the charge carriers.\cite{7,8}

To probe the intrinsic properties of FeTe$_{1-x}$Se$_{x}$ without the influence of excess Fe, some previous works have been performed to remove the effect of excess Fe by annealing in different conditions.\cite{8,9,10,11,12,13,14} Among those, bulk superconductivity was proved to be induced by annealing with appropriate amount of O$_2$\cite{8,11} or Te\cite{13}. In this paper, we reported that bulk superconductivity in Fe$_{1+y}$Te$_{1-x}$Se$_{x}$ can also be induced by annealing in Se and S vapor. The well-annealed samples show a large critical current density, \emph{J}$_c$. Combined with our previous reports about O$_2$ and Te annealing effects,\cite{8,11,13} we proved that bulk superconductivity can be successfully induced in Fe$_{1+y}$Te$_{1-x}$Se$_{x}$ by annealing in atmosphere of all chalcogens, such as O, S, Se and Te.

Single crystal with a nominal composition FeTe$_{0.6}$Se$_{0.4}$ was grown by the self-flux method as reported before.\cite{11} For annealing, the obtained crystals were cut and cleaved into thin slices with dimensions about 2.0 $\times$ 1.0 $\times$ 0.03 mm$^3$, weighed and loaded into a well-baked quartz tube (\emph{d} $\sim$ 10 mm $\phi$) separately with appropriate amount of Se or S grains. The quartz tube was carefully evacuated to the pressure better than 10$^{- 2}$ Torr, and sealed into the length of $\sim$ 100 mm. Then the crystals were annealed at a fixed condition, 400 $^{\circ}$C for 20 h, followed by water quenching. The vapor pressure for Se and S at 400 $^{\circ}$C is larger than 1 and 100 Torr, respectively. Thus, in our annealing condition, both Se and S can vaporize easily. Magnetization measurements were performed using a commercial superconducting quantum interference device (SQUID).  Microstructural and compositional investigations of the sample were performed using a scanning electron microscope (SEM) equipped with an energy dispersive x-ray spectroscopy (EDX).
\begin{figure}\center

　　\includegraphics[width=8cm]{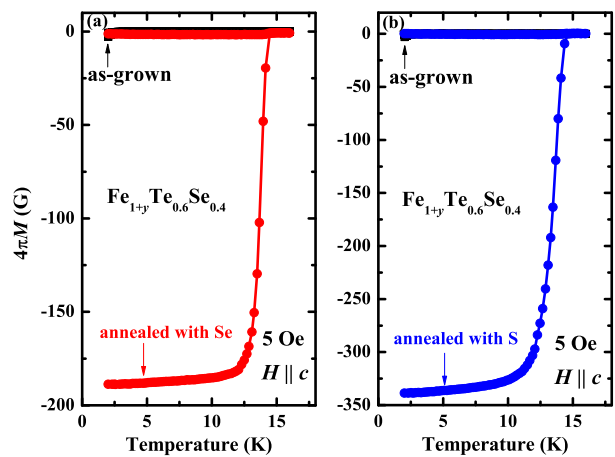}\\
　　\caption{(Color online) Temperature dependence of zero-field-cooled (ZFC) and field-cooled (FC) magnetization at 5 Oe for Fe$_{1+y}$Te$_{0.6}$Se$_{0.4}$ single crystals annealed at 400 $^{\circ}$C in (a) Se or (2) S vapor.}\label{}
\end{figure}

As-grown Fe$_{1+y}$Te$_{0.6}$Se$_{0.4}$ single crystal usually shows no superconductivity or very weak diamagnetic signal below 3 K. After annealing in Se or S vapor, superconductivity can be successfully induced. Similar to the case of annealing in Te vapor,\cite{13} the best quality sample can be obtained when the molar ratio of Se/S to the sample is $\sim$ 0.1. Temperature dependence of zero-field-cooled (ZFC) and field-cooled (FC) magnetization at 5 Oe for the samples annealed in Se or S vapor is shown in Figure 1 (a) and (b), respectively. The Se/S annealed samples show \emph{T}$_c$ higher than 14 K with a sharp transition width $\sim$ 1 K (obtained from the criteria of 10 and 90\% of the ZFC magnetization at 2 K). The high \emph{T}$_c$ and narrow transition width manifest the quality of the annealed samples is very high.

\begin{figure}\center

　　\includegraphics[width=8cm]{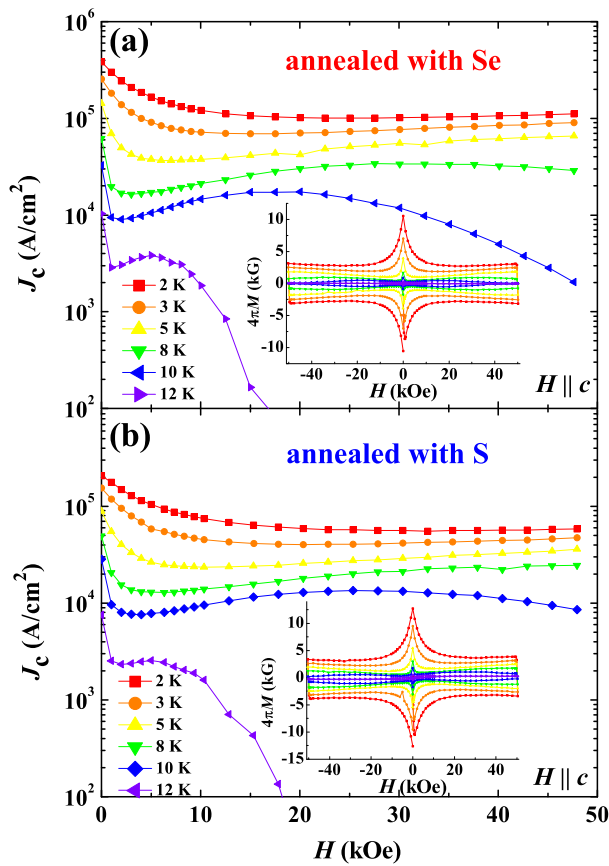}\\
　　\caption{(Color online) Magnetic field dependence of critical current densities of (a) Se and (b) S annealed Fe$_{1+y}$Te$_{0.6}$Se$_{0.4}$ for \emph{H} $\|$ \emph{c} at temperatures ranging from 2 to 12 K. Insets are the magnetic hysteresis loops for the annealed samples}\label{}
\end{figure}
To further confirm the superconducting properties of the samples annealed in Se/S vapor, magnetic hysteresis loops (MHLs) were measured and shown in the insets of Figure 2 (a) and (b), respectively. A second magnetization peak (SMP) can be witnessed, which is a common feature of IBSs. From the MHLs, we can obtain \emph{J}$_c$ by using the Bean model\cite{15}: \emph{J}$_c$ = 20$\Delta$\emph{M}/\emph{a}(1-\emph{a}/3\emph{b}), where $\Delta$\emph{M} is \emph{M}$_{down}$ - \emph{M}$_{up}$, \emph{M}$_{up}$ and \emph{M}$_{down}$ are the magnetizations when sweeping fields up and down, respectively, \emph{a} and \emph{b} are sample widths (\emph{a} $<$ \emph{b}). As shown in Figure 2, after being well annealed in Se or S vapor, the sample manifests a relatively high value of \emph{J}$_c$ $\sim$ 2 - 4 $\times$ 10$^5$ A/cm$^2$ under zero field, which is one of the largest among those reported in Fe(Te,Se) bulk samples.\cite{8,9,11,13} Besides, the \emph{J}$_c$ is also robust under applied field at low temperatures. The large value of \emph{J}$_c$ indicates that the superconductivity induced by annealing in Se/S vapor is in bulk nature, and this technique is promising to be used in future to enhance the superconducting properties of iron chalcogenide wires and tapes.

Figure 3(a) and (b)  show the SEM images of the crystals well annealed in Se and S vapor, respectively. After annealing, surface layers of the single crystal turn into polycrystal-like, which can be seen more clearly in the enlarged images shown in the insets. EDX measurements show that the polycrystal-like surface for crystal annealed in Se/S contains only Fe and Se/S with a molar ratio roughly 1 : 1. X-ray diffraction measurements show that the surface layers are most probably Fe$_7$Se$_8$ and FeS for samples annealed in Se and S vapor, respectively (data not shown). Those results are very similar to the case of sample annealed in Te vapor, which also shows a polycrystal-like surface composed of FeTe$_2$. After cleaving the polycrystal-like surface layers, the inner parts of the sample still keeps mirror-like surface, and EDX shows that the inner part is composed of Fe$_{1+y}$Te$_{0.6}$Se$_{0.4}$.

Compiling the above results with our previous reports,\cite{8,11,13} it is very interesting to find that for iron-chalcogenide Fe$_{1+y}$Te$_{1-x}$Se$_{x}$, superconductivity can be successfully induced by annealing in atmosphere of all chalcogenide elements, such as O, S, Se and Te (Polonium is not tested because of its radioactive nature). And after annealing, the surface layers turn to iron chalcogenide binary compounds Fe\emph{A}$_\emph{x}$ (\emph{A} : O, S, Se, Te). Actually, in iron chalcogenides, the excess Fe is almost unavoidable during the fabricating process, which is strongly magnetic and acts as a paring breaker.\cite{6} Annealing in chalcogenide atmosphere can induce superconductivity by removing the excess Fe in the crystal, forming stable Fe\emph{A}$_\emph{x}$ compounds on the surface. Based on this observation, we can speculate that any active elements, which can supply vapor to form stable binary compounds with excess Fe and is not harmful to the crystal itself may be able to induce superconductivity in Fe$_{1+y}$Te$_{1-x}$Se$_{x}$. Furthermore, this annealing method can easily provide high-quality samples with large value of \emph{J}$_c$, which is promising to be applied in fabricating iron chalcogenide wires and tapes. \begin{figure}\center

　　\includegraphics[width=8cm]{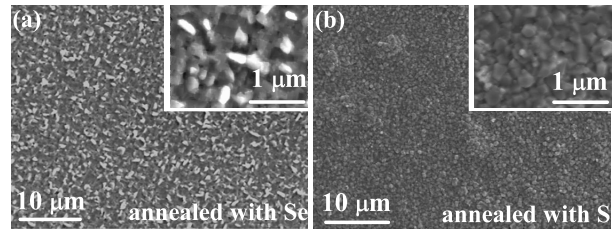}\\
　　\caption{(a) Scanning electron microscope (SEM) images for (a) Se annealed, (b) S annealed Fe$_{1+y}$Te$_{0.6}$Se$_{0.4}$ single crystal. Insets are enlarged images.}\label{}
\end{figure}

In summary, we found that bulk superconductivity with \emph{T}$_c$ above 14 K, can be induced in Fe$_{1+y}$Te$_{1-x}$Se$_{x}$ by annealing in Se and S vapor. The annealed sample shows a large value of \emph{J}$_c$ $\sim$ 2 - 4 $\times$ 10$^5$ A/cm$^2$.  Combined with our previous reports on O$_2$ and Te annealings, we proved that bulk superconductivity can be induced by annealing in atmosphere of all chalcogenide elements.

\begin{acknowledgment}

This work was partly supported by the Natural Science Foundation of China, the Ministry of Science and Technology of China (973 project: No. 2011CBA00105), and the Japan-China Bilateral Joint Research Project by the Japan Society for the Promotion of Science.
\end{acknowledgment}

$^{*}$zxshi@seu.edu.cn   $^{\dag}$tamegai@ap.t.u-tokyo.ac.jp

\end{document}